\documentclass[amsmath,twocolumn,amssymb,prd,preprintnumbers,showpacs]{revtex4}
\usepackage{amsmath}
\def \be {\begin{equation}}
\def \ee {\end{equation}}
\def \bea {\begin{eqnarray}}
\def \eea {\end{eqnarray}}
\usepackage{graphicx}
\usepackage{amssymb}
\begin{document}
\title{Causality and stability for Lorentz-CPT violating 
electrodynamics with dimension-5 operators}
\author{C. Marat Reyes}
\email[Electronic mail: ]{carlos.reyesm@usach.cl}
\affiliation{Departamento de F{\'i}sica, Universidad Aut{\'o}noma 
Metropolitana Iztapalapa, San Rafael Atlixco 186, C.P. 09340, 
M\'exico D.F., M\'exico, \\and
 Departamento de F{\'i}sica, 
 Universidad de Santiago de Chile, Casilla 307, Santiago, Chile}
\date{December 2010; 
published as Phys.\ Rev.\ D {\bf 82}, 125036 (2010)}
\begin{abstract}
Stringent limits on the Myers-Pospelov timelike parameter for photons 
$\xi<10^{-15}$ coming from astrophysical tests suggest 
exploring more general preferred
backgrounds, such as spacelike and lightlike. 
We take some steps in this direction. We allow 
the external four vector $n$ characterizing
the Lorentz symmetry breaking to have arbitrary 
directions in spacetime.
We study the consistency for the effective field theories
in each privileged frame by demanding
causality, stability and analyticity.
When specializing to a timelike background we found
runaway solutions leading to causality and unitarity violations.  
We show that the lightlike theory is a  
higher-derivative theory with more degrees of freedom and 
nonanalytic solutions leading to instabilities
when interactions 
are turned on.
We demonstrate by explicit calculation that both stability 
and analyticity are
preserved for the purely spacelike case while microcausality is highly suppressed.
This new anisotropic model 
opens the possibility to play a role in the search for
Planck-scale effects.
\end{abstract}
\maketitle
\section{Introduction}
The possibility of Lorentz and CPT symmetry violation  
has been actively explored both theoretically and experimentally 
in the past recent years.
Much of the motivation for considering such possibility come 
from the vision that spacetime at the Planck scales may  
depart drastically from its continuum description leaving some evidence at 
low energies, idea that has been reinforced by several candidate 
fundamental theories.
The possible observability in a form of
Lorentz violation has also been center of many interests for providing a   
route to realistic experimental scenarios where to test fundamental physics.

Searches for Lorentz and CPT violation  
have been performed in the following contexts:
string theory 
\cite{strings,strings1}, standard model extension \cite{sme}, 
spacetime foam \cite{foam},
loop quantum gravity \cite{lqg}, non commutative geometry \cite{ncg}, 
modified dispersion relations \cite{mdr}, 
cosmologically varying scalars \cite{Ralf1}, higher derivative 
field theory \cite{higher-derivative,higher-derivative2}, 
ultraviolet regulators \cite{regulator}, 
gravity \cite{gravity}, nonlinear electrodynamics \cite{vev}. 
The range of their predictions includes all matter sectors and gravity 
and have been probed by a large number 
of experimental tests, see the tables in Ref. \cite{Rusell-K}.
At present the best constraints, coming from
astrophysical observational tests, are imposed on photons 
which also violate discrete CPT symmetry. 
For example, from vacuum birefringence 
one has the bound  
$k_{AF}^{(3)}<2\times 10^{-42}$ GeV for   
the Chern-Simons parameter \cite{c.s,kostelecky-mewes1} 
and from ultra high energy cosmic rays $\xi<10^{-15}$ for the 
timelike parameter in the Myers and Pospelov model \cite{Gala-Sigl,Maccione:2007yc}.
Given such small numbers they are hardly 
justified for Lorentz violating corrections induced from the Planck-scale
and therefore 
for all phenomenological purposes they can be set to zero. 
From this consideration, it has been suggested to 
extend the Myers and Pospelov model to 
include more general backgrounds \cite{spacelike}.

In the present work we study Lorentz violations 
in the framework of effective field theory with the extra ingredient to be incorporated by means of
higher dimensional operators.
Typically one has a modified quadratic
Lagrangian density where some Lorentz violating background tensors 
are 
contracted with an operator.
As formulated in the standard model extension \cite{sme}, 
these background tensors can be viewed to arise from  
spontaneous symmetry breaking taking place in an underlying theory.
Higher dimensional operators have become more popular 
in the description of Lorentz violation in the last years. 
One example is the 
Myers and Pospelov model where the Lorentz violation 
is incorporated via a fixed four vector pointing only in the 
time direction and coupled to a dimension-5 operator. 
It has produced bounds 
from astrophysical observations \cite{Gala-Sigl,Maccione:2007yc}, 
synchrotron radiation \cite{urrutia}, radiative corrections \cite{reyes,Mariz:2010fm}, 
and laboratory tests \cite{b-a}.
Recently an extension for the photon sector including arbitrary mass 
dimension operators has also been constructed \cite{kostelecky-mewes1}. 

The incorporation of higher dimensional operators 
present some challenges and some caution has to be taken. 
These theories have been questioned for different 
reasons and many of the problems can be traced back to the bottomless of the Hamiltonian. 
That is, usually the
theory has an unbound negative energy part which under 
interactions can couple leading to instabilities.
On other hand, solutions of the plane wave ansatz can depend non 
analytically on the perturbative parameter, therefore 
undermining the validity of the effective theory. 
The use of higher dimensional operators not 
necessarily represent 
these problems, but a comprehensive study under 
which conditions the effective theory is consistent needs to be performed.
In this work we take some steps in this direction. 
We perform a systematic study of the 
causal, stability and analytical behavior
of the effective theories that introduce 
dimension-5 operators. 
We focus on the Myers-Pospelov 
modified theory of electrodynamics
allowing to consider an arbitrary symmetry breaking four vector. 

The outline of this paper is as follows.
In Sec II, we introduce the Myers-Pospelov model of 
electrodynamics with an arbitrary 
spontaneous symmetry breakdown direction. The field equations 
 and the dispersion relation are obtained.
In Sec III, we analyze the propagation properties for 
the isotropic and anisotropic cases, and we emphasize the nonanalyticity  
and instabilities for the lightlike case.
In Sec IV, we derive the retarded Green functions 
in the Coulomb gauge. 
Sec V is concerned with the quantum field theory, 
we obtain the propagator in the 
Lorentz gauge and we study microcausality for the purely spacelike case.
\section{Myers-Pospelov electrodynamics in arbitrary backgrounds}
Consider the Maxwell action
in the presence of a current $j_{\mu}$ plus the Myers and Pospelov term
\begin{eqnarray}\label{M-P.LAGRANGIAN}
S&=&\int d^4x\left(-\frac{1}{4}F_{\mu \nu}F^{\mu 
\nu}-4\pi j_{\mu}A^{\mu}  \right. \\ &&\left.+\frac{\xi}{M_P} n^{\mu}F_{\mu \nu}(n 
\cdot \partial) n_{\alpha} \widetilde F^{\alpha \nu}   \right),\nonumber 
\end{eqnarray}
where $n$ is an external four vector  
defining the type of background and fixed once for all, $\xi$ is 
a dimensionless parameter, $M_P$ the Planck mass,
$F_{\mu \nu }$ the electromagnetic field strength tensor and
$\widetilde F^{\alpha \beta}=\frac{1}{2}\epsilon^{\alpha \beta \rho 
\lambda}F_{\rho \lambda}$ its dual \cite{higher-derivative}. 

In its original proposal by Myers and Pospelov the action has the form
\begin{eqnarray}
S_{(5)}=\frac{\xi}{M_P}\int d^4x  \epsilon^{ijk} 
\dot A_i    \partial_j \dot  A_k,
\end{eqnarray}
resulting from choosing the external four vector 
in the purely timelike direction \cite{higher-derivative}.
This action and its corresponding dispersion relation 
has been the starting point for numerous phenomenological 
searches \cite{Maccione:2007yc}. In the following we will
extend the above treatment to include more general backgrounds
thus 
considering the general cases for $n$ 
incorporating the spacelike and lightlike cases.

Consider
the full Myers-Pospelov action re-expressed as
\begin{eqnarray}
S_{(5)}=-\frac{g}{2}\int d^4x  \epsilon^{\mu\nu \lambda \sigma} n_{\mu}
A_{\nu} (n \cdot \partial)^2  F_{\lambda \sigma},
\end{eqnarray}
where we have defined $g=\xi/M_P$. 
We observe the resemblance of this expression with 
the CPT-odd Chern Simons 
term \cite{c.s,KLINKHAMER.C.S,Andrianov}. This can 
be better seen by using the notation
of the recent standard model extension including nonrenormalizable 
operators \cite{kostelecky-mewes1}, where one has 
\begin{eqnarray}
(k_{AF}^{(5)})_{\mu}=-g(n\cdot \partial)^2n_{\mu}, 
\end{eqnarray}
\begin{eqnarray}
 (k^{(3)}_{AF})_{\mu}=m 
n_{\mu}/2,
\end{eqnarray}
for the Myers-Pospelov and Chern-Simons modifications 
respectively. It can be useful to
note that the Myers and Pospelov parameter is 
obtained from the replacement $m\to -2g(n \cdot \partial)^2 $.

The field equations for the action (\ref{M-P.LAGRANGIAN}) are
\begin{eqnarray}
\partial_{\mu} F^{\mu \nu}+g \epsilon ^{\nu \alpha 
\lambda \sigma} n_{\alpha}  (n 
\cdot \partial)^2   F_{\lambda \sigma}=4\pi j^{\nu},
\end{eqnarray}
which can be rewritten as
\begin{eqnarray}
\partial_{\mu} G^{\mu \nu}=4\pi j^{\nu},
\end{eqnarray}
where 
\begin{eqnarray}
G^{\mu \nu}= F^{\mu \nu}+2g  \epsilon ^{\mu \nu \alpha \beta}
n_{\alpha}  (n \cdot \partial)^2    A_{\beta}.
\end{eqnarray}
By introducing the notation $A^{\mu}=(A^0,{\bf A}=A^i)$, 
and the conventions $\eta_{\mu \nu}=\mathrm {diag} (1,-1,-1,-1)$,
$\epsilon^{0123}=\epsilon^{123}=1$ we can write the 
electric and magnetic fields as
\begin{eqnarray}
F_{0i}=E^i, \qquad F_{ij}=-\epsilon^{ijk}B^ k,
\end{eqnarray}
or in vectorial notation 
\begin{eqnarray}\label{ELECTRICFIELD}
{\bf E}=-\frac{\partial {\bf A}}{\partial t} - \nabla A_0,
\end{eqnarray} 
\begin{eqnarray}
{\bf B}= \nabla \times {\bf A}.
\end{eqnarray} 
In terms of the physical fields for a general 
$n=(n_0,{\bf n} )$ we have the field equations
\begin{eqnarray}\label{ELECTREC1}
\nabla \cdot {\bf E}+2g(n\cdot \partial)^2({\bf n} \cdot {\bf B})=4\pi \rho,
\end{eqnarray}
\begin{eqnarray}
-\frac{\partial {\bf E}}{\partial t}+\nabla \times {\bf B}+
2g (n\cdot \partial)^2 (n_0 {\bf B}-({\bf n} 
\times {\bf E}))=4\pi {\bf j}, \label{ELECTREC}
\end{eqnarray}
together with the usual homogeneous ones
\begin{eqnarray}\label{homoge}
  \nabla \cdot {\bf B}=0, \qquad \nabla 
  \times {\bf E}+ \frac{\partial 
  {\bf B}}{\partial t}=0. 
\end{eqnarray}
Using the equations (\ref{ELECTREC}) and the last of (\ref{homoge})
and in the absence of sources, we have 
\begin{eqnarray}
 \Box {\bf E} &+&\nabla ( \nabla \cdot {\bf E})+
2g (n\cdot \partial)^2  \nonumber \\ &&
\times \left(n_0 (\nabla \times 
{\bf E})+\frac{\partial}{\partial t}
({\bf n} \times {\bf E})\right)= 0.
\end{eqnarray}
In order to find the dispersion relation
we pass to momentum space by considering the ansatz ${\bf E}(x)= 
\widetilde {\bf E}(k) e^{-i k\cdot x}$. Replacing above, we obtain
\begin{eqnarray}
 k^2 \widetilde {\bf E} &+&{\bf k}( {\bf k} \cdot \widetilde {\bf E})+
2ig  (n\cdot k)^2 \nonumber \\ &&
\times  (n_0 ({\bf k} \times \widetilde {\bf E})-k_0
({\bf n} \times \widetilde {\bf E}))= 0,
\end{eqnarray}
and solving for the determinant gives us the Myers 
and Pospelov covariant dispersion relation 
\begin{eqnarray}\label{DISP.RELATION}
(k^2)^2-4g^2(n \cdot k)^4 \left((k\cdot n)^2-k^2n^2\right)=0,
\end{eqnarray}
which will be crucial for the rest of the work.
\section{Dispersion relations and energy stability} \label{SECTION2}
In this section we study the solutions to the dispersion relation (\ref{DISP.RELATION})
when $n$ is chosen to be purely timelike, purely spacelike and lightlike. 
For each of the resulting effective theories we carry out an analysis
over the necessary conditions for stability, 
causality and analyticity.
\subsection{An additional criteria: Analyticity}
Effective field theories that incorporate higher dimensional operators to describe 
perturbative corrections to
a conventional field theory may present some drawbacks. 
In particular those that include dimension-5 operators as the ones we are interested in here.
First, the higher-derivative theory may have 
unbounded energy
from below leading to instabilities problems 
when interactions come into play.
Second, even the slightest inclusion of 
higher-derivative terms 
can produce an 
increase in the degrees of freedom with respect to the 
unperturbed theory. 
Both the negative energies and the 
increase of degrees of freedom are somehow 
related. 
It turns out that
in general the degrees of freedom that have been incremented are also 
responsible for the appearance of negative energies. 
Moreover, this
class of solutions have 
nonanalytical behavior in the perturbative parameter, that is, they 
tend to infinity when the perturbative parameters are taken to zero.
Some perturbative methods have been developed in order to eliminate 
these additional degrees of freedom. The resulting effective theories are shown to have
positive and hermitian Hamiltonians \cite{pert}.
In the next subsection, we consider the analytical criteria also called  
perturbative constraint \cite{pert2} in order to discriminate 
whether a theory is a higher derivative theory or not. 
\subsection{The isotropic model}
\begin{figure} \label{Fig1}
\centering
\includegraphics[
width=0.5\textwidth]{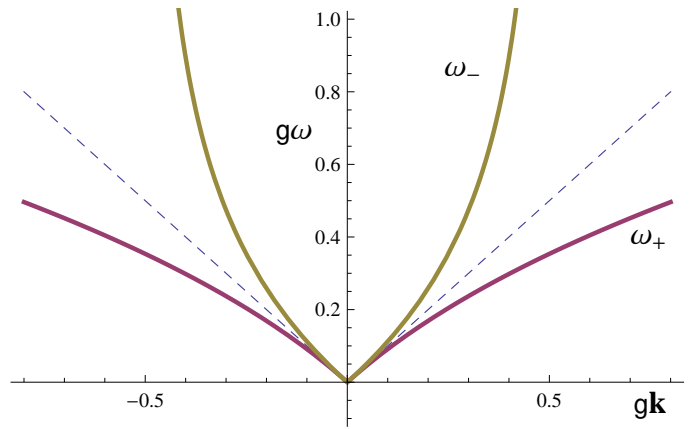}
\caption{The dispersion relation (\ref{disp.rel-timelike}) 
for a purely timelike four vector $n=(1,0,0,0)$ with corresponding curves $\omega_{-}$ and $\omega_{+}$ 
and the dashed line corresponding to the  
light cone.}
\end{figure}
We start considering the isotropic four-vector 
$n=(n_0,0,0,0)$ for which the dispersion relation (\ref{DISP.RELATION})
has the form 
\begin{eqnarray}
(k ^2)^2-4g^2 k_0^4 {\bf k}^2 n_0^6=0.
\end{eqnarray}
Solving we obtain the frequency solutions 
\begin{eqnarray}\label{disp.rel-timelike}
\omega_{\lambda}({\bf k})=\frac{{\left|\bf k\right|} }
{\sqrt{1+2g \lambda {\left|\bf k\right|}n_0^3}},
\end{eqnarray}
with the two polarizations $\lambda=\pm$. 
The solutions correctly reproduces the usual ones in the limit $g \to 0$
and coincide with those previously derived in \cite{urrutia}.
 
For momenta
$\left|\bf k\right|<1/(2gn_0^3)$ the approximation of (\ref{disp.rel-timelike}) 
\begin{eqnarray}\label{dr-approx}
\omega_{\lambda}({\bf k})\approx\left|\bf k\right|-
 g\lambda n_0^3 \left|\bf k\right|^2,
\end{eqnarray}
gives the cubic modifications reported in \cite{higher-derivative}. 
For higher momenta the approximation is no longer valid for $\omega_{-}$
and its imaginary part leads to the loss of unitarity and to instabilities.
This signals the need of an ultraviolet cut-off 
function in order to avoid runaway solutions.
The quantization of 
this model restricted to a region of momenta 
$\left|\bf k\right|< \left|\bf k\right|_{\text{max}}=1/(2gn_0^3)$ 
and using
Pauli-Villars regularization methods has been performed in \cite{reyes}. 

To discuss stability and causality 
we plot the
dispersion relation (\ref{disp.rel-timelike}) in Fig.\ 1 and consider the 
necessary conditions discussed in \cite{ralf}.
For $k_{+}=(\omega_{+},{\bf k})$ we observe 
the possibility to have 
negative energy by performing an observer 
Lorentz transformation to a boosted frame
in which $\omega_{+}$ is negative. 
For $n_0=1$ and fixed momentum this occurs when 
$1/\sqrt{1+2g \left|\bf k\right|}<\left|\bf v\right|<1$ and together
with the maximum allowed momentum $\left|\bf k\right|_{\text{max}}$ this leads
to the requirement that the allowed concordant 
frames \cite{ralf} in 
which the quantization will remain consistent 
are such that the boost velocity restrict to $\beta< 1/\sqrt {2}$
with respect to the rest frame.

Consider the group velocity
\begin{eqnarray}
v_{g^\pm}({\bf k})&=&\frac{ (1\pm g  {\left|\bf k\right|} n_0^3  
)}{(1\pm2g {\left|\bf k\right|} n_0^3   )^{3/2}},
\end{eqnarray}
and note that $v_{g^{-}}({\bf k})$ can exceed the speed 
of light introducing problems of causality.
In the next sections we compute the retarded Green function 
which gives 
us a more transparent way to analyze causality violations in 
the presence of interactions.
\subsection{The spatially anisotropic model}
For a generic external four vector $n=(n_0,{\bf n})$
the dispersion relation (\ref{DISP.RELATION})
will be a sixth order polynomial with 
some roots having imaginary parts. 
Also, the high order character of the equation 
makes technically difficult
to obtain the exact solutions.
To simplify we 
will restrict to the type of anisotropies 
introduced by lightlike and purely spacelike preferred vectors.

Consider first the lightlike case $n=(n_0,{\bf n})$ with $n^2=0$,
for which the dispersion relation is
\begin{eqnarray}
k_0^2-{\bf k}^2 +2g\lambda \left(n_0k_0-\left|\bf k\right| 
 \left|\bf n\right|\cos\theta \right)^3=0.
\end{eqnarray}
The solutions are
\vspace{-20pt}
\begin{widetext}
\begin{eqnarray}\label{sol-lightlike}
\omega_{1}(\lambda,{\bf k})&=&\frac{\lambda}{6g  
\left|\bf n\right|^3 }\left(  6g\lambda 
z\cos\theta-1+\frac{(1-12g\lambda 
z\cos\theta )}{\Delta^{1/3} }
+\Delta^{1/3}\right), \nonumber \\
\omega_{2} (\lambda,{\bf k})&=&\frac{\lambda}
{6g  \left|\bf n\right|^3 }\left(6g\lambda 
z\cos\theta-1-
\frac{(1+i\sqrt{3})(1-12g\lambda z\cos\theta )}
{2 \Delta^{1/3} } -\frac{(1-i\sqrt{3})
\Delta^{1/3}}{ 2}\right),  \\
\omega_{3} (\lambda,{\bf k})&=&\frac{\lambda}{6g  
\left|\bf n\right|^3 }\left(6g\lambda 
z\cos\theta-1-
\frac{(1-i\sqrt{3})(1-12g\lambda z\cos\theta )}
{2 \Delta^{1/3}} -\frac{(1+i\sqrt{3})
\Delta^{1/3}}{ 2}\right),\nonumber
\end{eqnarray}
\end{widetext}
where 
\begin{eqnarray}
\Delta(\lambda,z )&=&\Delta_1(\lambda,z )+\Delta_2(\lambda,z ),
\end{eqnarray}
and
\begin{eqnarray}
\Delta_1(\lambda,z )&=&-1+54g^2 z^2
+18g\lambda z\cos\theta-
54g^2z^2  \cos^2\theta,\nonumber \\ \noindent
\\
\Delta_2(\lambda,z )&=&(\left(-1+12g\lambda 
z\cos\theta \right)^3 +(1-18
g\lambda z\cos\theta  \nonumber \\
&& -27g^2z^2\left(1-2\cos 
\left(2\theta \right)\right)  )^2)^{1/2}.
\end{eqnarray}
Above we 
have made explicit the relation $n^2=0$, defined 
$z=\left|\bf k\right| \left|\bf n\right|^3$
and considered $\theta$ the angle between ${\bf k}$ and ${\bf n}$. 

Consider the approximations of small $g$ at linear 
order:
\begin{eqnarray}
\omega_{1,3}(\lambda,{\bf k})&\approx& \lambda 
\left|\bf k\right|+\mathcal O(g), \nonumber 
\\
\omega_{2}(\lambda,{\bf k})&\approx& \frac{\lambda }
{2 g \left|\bf n\right|^3}+3 \left|\bf k\right| 
\cos \theta+\mathcal O(g).
\end{eqnarray}
We have two solutions that approximate well the usual solutions when taking the limit $g\to 0$
and one extra degree of freedom that is nonanalytic in the perturbative parameter.
We expect the lightlike theory to develop 
negative energies and have 
nonunitarity evolution.

For the case where $n$ is purely spatial 
$n=(0,{\bf n})$ we have
\begin{eqnarray}\label{spacelikecase}
(k^2)^2-4g^2 ({\bf k}\cdot {\bf n})^4  
\left(({\bf k}\cdot {\bf n})^2 +(k_0^2-
{\bf k}^2){\bf n}^2\right)=0,
\end{eqnarray}
with roots 
\begin{eqnarray}\label{disp.rel-spacelike}
\omega_{\pm}({\bf k})&=&({\bf k} ^2+2g^2{\bf n}^2 
({\bf k}\cdot {\bf n})^4 
 \nonumber  \\
 && \pm 2g \left|{\bf k}\cdot {\bf n}\right|^3  
 \sqrt{ 1+g^2  {\bf n}^4
 ({\bf k}\cdot {\bf n})^2 })^{1/2}.
\end{eqnarray}
For small momenta we have at leading order
\begin{eqnarray}
\omega_{\lambda}({\bf k})\approx\left|\bf k\right|-
 g\lambda \left|{\bf n}\right|^3  \left|\cos 
 \theta \right|^3  \left|\bf k\right|^2,
\end{eqnarray}
where again $\theta$
is the angle between ${\bf k}$ and ${\bf n}$.
The above expression is very similar to the one given 
in Eq (\ref{dr-approx}), but valid for all momenta. 

The group and phase velocities are, respectively 
\vspace{-30pt}
\begin{widetext}
\begin{eqnarray}
{ v}_{g^ \pm}=\frac{ 1+4g^2 
{\left|\bf k\right|}^2\cos^4\theta
  \pm g{\left|\bf k\right|} \left|\cos\theta\right|^3 (3+4g^2 
  {\left|\bf k\right|}^2\cos^2 \theta ) / \sqrt{1+g^2 
  {\left|\bf k\right|}^2 \cos^2\theta} }
{ (1+2g^2\left|\bf k\right|^2 
\cos^4\theta  \pm 2g {\left|\bf k\right|}\left|\cos\theta\right|^3
\sqrt{1+g^2 {\left|\bf k\right|}^2 \cos^2\theta}  )^{1/2}},  
\end{eqnarray}
\end{widetext}
and
\begin{eqnarray}
{ v}_{ph^ \pm}&=&(1+2g^2 {{\left|\bf k\right|}^2}\cos^4\theta 
\nonumber \\ && \pm 2g{{\left|\bf k\right|}}  \left|\cos\theta\right|^3 \sqrt{ 1+g^2 
 {{\left|\bf k\right|}^2} \cos^2\theta})^{1/2}, 
\end{eqnarray}
where we have set ${\bf n}^2=1$. 

We can 
distinguish three cases for the dispersion relation (\ref{disp.rel-spacelike}): 

(a) when $ {\bf k}$ and $ {\bf n}$ are perpendicular,

(b) parallel or anti-parallel with $\left|\cos \theta\right|=1$,

(c) directions such that $ \left| \cos \theta \right|<1$.
\begin{figure}
\centering
\includegraphics[
width=0.5\textwidth]{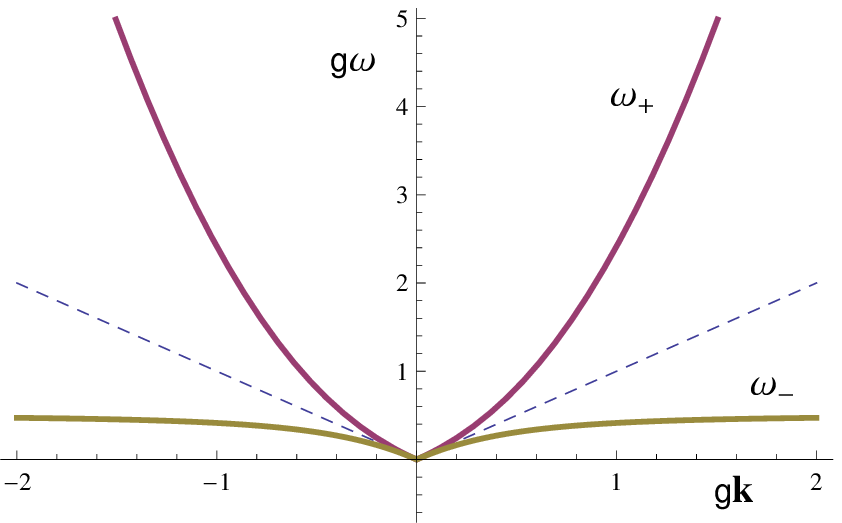}\label{fig2}
\caption{Dispersion relation (\ref{disp.rel-spacelike}) for the case (b) with ${\bf n}^2=1$
with corresponding curves $\omega_{+}$ and $\omega_{-}$ and the light cone (dashed).}
\end{figure}
\begin{figure}
\centering
\includegraphics[
width=0.5\textwidth]{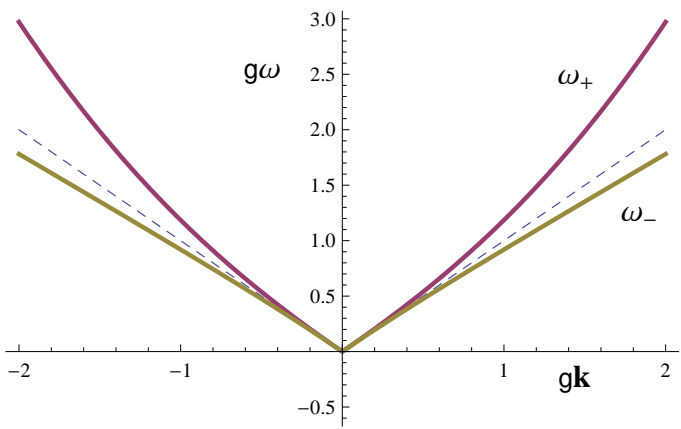}
\caption{Dispersion relation (\ref{disp.rel-spacelike}) for the case 
(c) with $\cos \theta =1/2$ and ${\bf n}^2=1$ with corresponding curves $\omega_{+}$ and $\omega_{-}$ 
and the light cone (dashed).}
\end{figure}
For the case (a)
we recover the usual dispersion relation $\omega=\left|\bf k\right|$, 
also mentioned in \cite{spacelike} as a blind direction.
To analyze the cases (b) and (c) consider the plots Fig.\ 2 and Fig.\ 3.

For the cases (b) and (c) the momentum
$k_{-}=(\omega_{-},k)$ is outside the light cone which 
threatens the stability of the theory. 
Therefore for small momentum we must restrict to the region
\begin{eqnarray}
\beta_{\lambda}< 1- \lambda g\left|\bf k\right| \left|\cos\theta\right|^3. 
\end{eqnarray}
in order to 
have concordant frames.
This includes a large number of concordant frames since the 
boost velocity is very near the light cone.
Going to large momenta we have that $\omega_{-}$ approaches the constant value
$\omega_{-}({\bf k})\approx \frac{1}{2g}$ for the case (b)
and it diverges as
$\omega_{-}({\bf k})\approx \sin\theta \left|{\bf k}\right|$ for the case (c) 
and therefore the velocity group approaches the constant value
$v_{g^-}({\bf k})\approx \sin\theta$.
In consequence any finite boost will introduce instabilities  
in the parallel or anti-parallel directions for high momentum.
We expect to have more concordant frames for propagations 
approaching the perpendicular direction.

On the other hand we see that 
$v_{g^+}({\bf k})$ can exceed the speed of light which indicates possible
violations of microcausality.
\section{Causality and retarded Green functions}
In this section to study 
the causal behavior of the gauge fields in the presence of interactions
we compute the retarded Green function.
As standard we assume a delta source interaction and to 
avoid to deal with non-physical degrees of freedom
we work in the Coulomb gauge $\nabla \cdot {\bf A}=0$.
\subsection{Transverse Green functions}
The transverse Green function will be constructed by first reducing 
the degree of freedom corresponding to $A_0$. 
Hence, from (\ref{ELECTRICFIELD}) and (\ref{ELECTREC1}) 
we replace the non dynamical relation 
\begin{eqnarray}
A_0=2g\frac{ (n \cdot \partial)^2  }{ \nabla^2}({\bf n}\cdot  
 {\bf B})-
4\pi \frac{1}{ \nabla^2}\rho,
\end{eqnarray}
in (\ref{ELECTREC}) to yield the expression
\begin{eqnarray}\label{eqgauge}
 && \Box {\bf A}+
2g \frac{(n\cdot \partial)^2}{\nabla^2}
\nabla \left({\bf n} \cdot \nabla \times \frac{\partial {\bf A}}
{\partial t} \right)+
2g (n\cdot \partial)^2  \nonumber \\
&& \times \left(n_0  \nabla \times {\bf A} + {\bf n} 
\times \frac{\partial {\bf A}}{\partial t}\right)
\nonumber \\ &&
+ 4g^2 \frac{(n\cdot \partial)^4}{\nabla^2} {\bf n} \times   
\nabla \Big({\bf n }\cdot   \nabla \times 
{\bf A}\Big)=4 \pi {\bf  J}_T.
\end{eqnarray}
Above we have used current conservation 
$\partial \rho/\partial t+\nabla \cdot {\bf j}=0$
and defined the transverse current ${\bf J}_T$ 
\begin{eqnarray}\label{JT}
 {\bf  J}_T={\bf j}- \frac{ \nabla ( \nabla \cdot 
 {\bf j})}{ \nabla^2}+2g  
 \frac{ (n  \cdot \partial)^2}{ \nabla^2}{\bf n} \times  \nabla \rho,
\end{eqnarray}
which can be easily checked to satisfy the conserved 
equation $\nabla \cdot {\bf J}_T=0$.
Now, provided the gauge field is transverse we have 
\begin{eqnarray} 
{\bf n} \times  \nabla ({\bf n }\cdot   \nabla \times {\bf A}) &=&
 ({\bf n}^2 \nabla^2-({\bf n}\cdot {\bf \nabla})^2) 
  {\bf A}\nonumber \\
  &&+ (({\bf n}\cdot 
  {\bf \nabla})\nabla-{\bf n}\nabla^2) ({\bf n}\cdot {\bf A}),\nonumber \\
\end{eqnarray}
which allows us to rewrite the last term in (\ref{eqgauge}) and so obtain 
\begin{eqnarray} \label{eq.of.motion}
 &&\left(\Box +4g^2 (n\cdot \partial)^4 
  \Big({\bf n}^2 -\frac{({\bf n}\cdot {\bf \nabla})^2}
  {\nabla^2}\Big)\right) {\bf A} +
2g \frac{(n\cdot \partial)^2}{\nabla^2}\nonumber \\
&&\times
 \nabla \left({\bf n} \cdot  \nabla \times \frac{\partial  {\bf A}}
{\partial t}  \right)+
2g (n\cdot \partial)^2   \left(n_0  \nabla \times {\bf A} + {\bf n} 
\times \frac{\partial {\bf A}}{\partial t}\right)\nonumber \\
&&
+4g^2 (n\cdot \partial)^4 \left(\frac{\nabla({\bf n}\cdot 
  {\bf \nabla})}{\nabla^2}-{\bf n}\right)({\bf n}\cdot {\bf A}) 
  = 4 \pi {\bf  J}_T.
\end{eqnarray}
Taking components of the above equation of motion we have
\begin{eqnarray} \label{eq.of.motion2}
M_{ij}(\partial_0,\nabla) { A}^j={ J}_T^i,
\end{eqnarray}
where
\begin{eqnarray} 
&&M_{ij}(\partial_0,\nabla) \nonumber \\
&=&\Big[\left(\Box +4g^2 (n\cdot \partial)^4 
  \Big({\bf n}^2 -\frac{({\bf n}\cdot {\bf \nabla})^2}
  {\nabla^2}\Big)\right) \delta_{ij} \nonumber \\&&+2 g (n\cdot \partial)^2  
\left( n_0    \epsilon^{ilj}      \partial_l 
+ \partial_0 \left(  \epsilon^{ilj}n^l+ 
\frac{\partial_i\partial_m}{\nabla^2} \epsilon^{lmj}n^l \right)\right)   
\nonumber \\
&& + 4g^2 (n\cdot  \partial)^4 \left(\frac{\partial_i
\partial_l n^l}{\nabla^2}-n^i \right)n^j \Big].
\end{eqnarray}
After some straightforward calculation 
the transverse retarded Green 
function is given by
\begin{eqnarray}
G_{jk}(x-y)=  T_{jk}(\partial_0,\nabla) \,D(x-y),
\end{eqnarray}
where the tensor
\begin{eqnarray} \label{transv-op}
  T_{jk}(\partial_0,\nabla)=\pi_{jk} \Box 
 +4g^2 (n\cdot  \partial)^4  s_{jk}+2g (n\cdot  \partial)^2 a_{jk},
\end{eqnarray}
is given in terms of the operators
\begin{eqnarray}\label{pi}
 \pi_{jk} =\left(\delta_{jk}-\frac{\partial_j\partial_k}{\nabla^2}\right),
\end{eqnarray}
\begin{eqnarray}
 s_{jk}=\left(n^j-\frac{({\bf n}\cdot \nabla)}{\nabla^2}\partial_j
  \right)\left(n^k-\frac{({\bf n}\cdot \nabla)}{\nabla^2}\partial_k \right),
\end{eqnarray}
\begin{eqnarray}
a_{jk}=\left(n_0+\frac{({\bf n}\cdot \nabla)}
 {\nabla^2}\partial_0\right)\epsilon^{jkl}\partial_l,
\end{eqnarray}
The scalar retarded 
Green function $D(x-y)$ is defined to satisfy 
\begin{eqnarray}
(\Box^2+4g^2(n \cdot \partial)^4 \left((  n \cdot 
\partial)^2-n^2\Box\right)) D(x-y)=4\pi\delta (x-y).\nonumber \\ 
\end{eqnarray}
The fourier transform  
lead us to the integral 
\begin{eqnarray}\label{causal-function}
&& D(x-y)=\frac{4\pi}{(2\pi)^4}\int_{C} d^4k \nonumber \\
&& \times\frac{e^{-ik\cdot (x-y)}}{(k^2)^2-4g^2(n \cdot k)^4 
 ((k\cdot n)^2-k^2n^2)},
\end{eqnarray}
where the contour of the curve $C$, in the complex $k_0$ plane, defines the 
boundary conditions to be imposed on the various Green functions.
For the retarded Green function 
the curve is defined above the real axis
and is denoted by $C_R$.
The corresponding scalar function $D_R(x-y)$ controls the causal behavior of the theory
and therefore to study causality will be enough to compute $D_R(x-y)$. 
\subsection{Causality for $n$ purely spacelike}
For the case $n=(0,{\bf n})$ the integral (\ref{causal-function}) is
\begin{eqnarray}
 D_R(z)=\frac{4\pi}{(2\pi)^4} \int d^3x e^{i{\bf k}\cdot 
 {\bf z}} \int_{C_R} dk_0  \frac{e^{-ik_0 z_0 }}{(k^2_0-\omega^2_{+})
 (k_0^2-\omega^2_{-})},\nonumber \\
\end{eqnarray}
where $z=x-y$ and the solutions $\omega_{\pm}$ are given by (\ref{disp.rel-spacelike}).

We observe that the integral 
\begin{eqnarray}
 I_R(z_0)= \int_{C_R} dk_0  \frac{e^{-ik_0 z_0}}{(k^2_0-\omega^2_{+})
 (k_0^2-\omega^2_{-})},
\end{eqnarray}
vanishes for $z_0<0$ since
the contour $C_R$ must be closed in the upper half plane therefore failing
to enclose any of the poles $\omega_{\pm}$ lying on the real axis.
We have in consequence that causality is preserved for the spacelike background.
\subsection{Causality for $n$ purely timelike}
Consider now $n=(1,0,0,0)$, we have from (\ref{causal-function})
\begin{eqnarray}\label{causalspace}
 D_R(x)&=&\frac{4\pi}{(2\pi)^4}\int d^3k \,
 \frac{e^{i{\bf k}\cdot 
 {\bf x}}}{(1-4g^2 \left|\bf k\right|^2)}  \nonumber \\
 &&\times \int_{C_R} dk_{0}  \frac{e^{-ik_0x_0}}{
 (k_0^2-\omega_{+}^2)(k_0^2-\omega_{-}^2)},
\end{eqnarray}
where from (\ref{disp.rel-timelike}) the solutions are
\begin{eqnarray}
\omega_{+}({\bf k})=\frac{{\left|\bf k\right|} }
{\sqrt{1+2g {\left|\bf k\right|}}},
 \qquad
\omega_{-}({\bf k})=\frac{{\left|\bf k\right|} }
{\sqrt{1-2g {\left|\bf k\right|}}},
\end{eqnarray}
and we have set $y=0$. 

The calculation of $ D_R(x)$ will be done in two stages depending on the sign of 
$x_0$.

(i) For $x_0<0$ we focus on the contour integral
\begin{eqnarray}\label{integral}
 I_R(x^{-}_0)=\int_{C_R} dk_{0}  \frac{e^{ik_0 \left|x_0\right|}}{
 (k_0^2-\omega_{+}^2)(k_0^2-\omega_{-}^2)}.
\end{eqnarray}
We must close the contour of $C_R$ in the upper half plane and therefore we have contributions 
only when when $\left|\bf k\right|>1/(2g)$ due to the pole in $i \left|\omega_{-}\right|$ with
\begin{eqnarray}
 \left|\omega_{-}\right|=\frac{{\left|\bf k\right|} }
{\sqrt{2g{\left|\bf k\right|}-1 } }.
\end{eqnarray}
The contour integral gives
\begin{eqnarray}\label{I-int}
I_R(x^ {-}_0)=\frac{-\pi e^{-\left|\omega_{-}\right| 
\left|x_0\right|}}
{\left|\omega_{-}\right|(\omega_{+}^2+\left|\omega_{-}\right|^2  )},
\end{eqnarray}
and from (\ref{causalspace}) we have
\begin{eqnarray}
 D_R(x^-)&=&\frac{1}{8\pi}\int_{1/(2g)}^{\infty} 
 d\left|\bf k\right| \nonumber \\
 &&\int_{-\pi}^{\pi}\sin\theta d\theta   \,
 \frac{e^{-\left|\omega_{-}\right| 
 \left|x_0\right|}}{ g\left|\bf k\right|\left|
 \omega_{-}\right|}e^{i\left|{\bf k}\right| 
 \left|{\bf x}\right|\cos\theta},
\end{eqnarray}
where we have used
\begin{eqnarray}
 \omega_{+}^2+\left|\omega_{-}\right|^2
  =\frac{-4g\left|\bf k\right|^3 }{1-4g^2\left|\bf k\right|^2}.
\end{eqnarray}
Integrating in the angle we arrive at
\begin{eqnarray}
D_R(x^{-})&=&\frac{1}{4\pi gr}\int_{1/(2g)}^{\infty}
d\left|\bf k\right|
\frac{e^{-\left|\omega_{-}\right| \left|x_0\right|}
 \sin(\left|\bf k\right|r )}{\left|\bf k\right|^2 
 \left|\bf \omega_{-}\right|},\nonumber \\
\end{eqnarray}
where we have introduced the notation $r=\left|{\bf x}\right|$.
In order to arrive to a more elegant 
expression we perform the change of variables $x=\frac{1}{\left|\bf k\right|}$
leading to
\begin{eqnarray}
D_R(x^ {-})&=&\frac{1}{4\pi gr}\int_{0}^{2g}
dx
 e^{-\frac{ \left|x_0\right|}{\sqrt{x(2g-x)}}}
 \sin \left(\frac{r}{x} \right) \nonumber \\
 &&\times \sqrt{x(2g-x)},
\end{eqnarray}
again making $z=x-g$ we have
\begin{eqnarray}
D_R(x^{-})&=&\frac{1}{2\pi gr}\int_{0}^{g}dz
e^{-\frac{ \left|x_0\right|}{\sqrt{g^2-z^2}}} 
\sqrt{g^2-z^2} \nonumber \\
&&\times
\cos \left(\frac{ rz }{g^2-z^2}\right) \sin 
\left(\frac{ gr}{g^2-z^2}\right). 
\end{eqnarray}
From the above expression one can already appreciate
 causality violations effects, however to 
end up with a closed expression for $D_R$ we will continue with the next case.

(ii) For $x_0>0$ the contour integral is
\begin{eqnarray}\label{integral2}
 I_R(x_0^{+})=\int_{C_R} dk_{0}  \frac{e^{-ik_0 \left|x_0\right|}}{
 (k_0^2-\omega_{+}^2)(k_0^2-\omega_{-}^2)},
\end{eqnarray}
which has to closed in the lower half plane.
For $\left|\bf k\right|<1/(2g)$ the integral includes the poles 
$\pm \omega_{+}$
and $\pm \omega_{-}$ and we obtain
\begin{eqnarray}\label{cont1}
 I_R(x^+_0)=-2\pi\left(\frac{\sin(\omega_{+}\left|x_0\right|)}
 {\omega_+(\omega_+^2-\omega_{-}^2)}-
  \frac{\sin(\omega_{-}\left|x_0\right|)}{\omega_{-}
  (\omega_+^2-\omega_{-}^2)}\right).
\end{eqnarray}
For $\left|\bf k\right|>1/(2g)$ 
we have the poles $-i\left|\bf \omega_{-}\right|$ and $\pm \omega_{+}$,
and we obtain
\begin{eqnarray}\label{cont2}
 I_R(x^{+}_0)&=&-2\pi\left(\frac{\sin(\omega_{+}\left|x_0\right|)}
 {\omega_+(\omega_+^2-\omega_{-}^2)} \right. \nonumber \\ &&\left.+ 
\frac{ e^{-\left|\omega_{-}\right| 
 \left|x_0\right|}}
 {2\left|\omega_{-}\right|(\omega_{+}^2+\left|\omega_{-}\right|^2  )
 }\right),
\end{eqnarray}
where we have introduced $\left|\omega_{-}\right|$ in the second term.

Considering the first terms in (\ref{cont1}) and (\ref{cont2}) we have a contribution
\begin{eqnarray}
D^ {(1)}_R(x^{+})&=&\frac{1}{2\pi gr}\int_{0}^{\infty}
d\left|\bf k\right|
\frac{ \sin(\omega_{+}\left|x_0\right|  )
 \sin(\left|\bf k\right|r )}{\left|\bf k\right|^2 
 \omega_{+}},\nonumber \\
\end{eqnarray}
and from the second term in (\ref{cont1}) we have
\begin{eqnarray}
D^ {(2)}_R(x^{+})&=&\frac{-1}{2\pi gr}\int_{0}^{1/(2g)}
d\left|\bf k\right|\nonumber \\ &&\times
\frac{ \sin(\omega_{-}\left|x_0\right|  )
 \sin(\left|\bf k\right|r )}{\left|\bf k\right|^2 
 \omega_{-}}.
\end{eqnarray}
The third contribution coming from the second term in (\ref{cont2}) is the same as 
the one in (\ref{I-int}) and therefore we have 
\begin{eqnarray}
D^ {(3)}_R(x^{+})=D_R(x^{-}).
\end{eqnarray}
Adding the three contributions we arrive at
\begin{eqnarray}
D_R(x^+)&=&\frac{1}{\pi gr} \left[   
\int_{g}^{\infty} dz 
 \sin(\frac{x_0}{\sqrt{z^2-g^2}}) \right. \nonumber \\ && \left.\times \sqrt{z^2-g^2}
 \cos\left(\frac{zr}{z^2-g^2}\right) \sin \left
 (\frac{gr}{z^2-g^2}\right) \right. \nonumber \\
 && \left.+\frac{1}{2}\int_0^g dz 
e^{\frac{ -\left|x_0\right|}{\sqrt{g^2-z^2}}}
 \sqrt{g^2-z^2} \right. \nonumber \\
 && \left.\times \cos\left(\frac{zr}{g^2-z^2}\right) 
 \sin \left(\frac{gr}{g^2-z^2}\right) \right].
\end{eqnarray}
The total scalar retarded Green function $D_R(x)=D_R(x^+)
\theta(x_0)+D_R(x^{-})\theta(-x_0)$ is given by
\begin{eqnarray}
D_R(x)&=&\frac{1}{\pi gr} \left[  \theta(x_0) \int_{g}^{\infty} 
dz  \sin(\frac{x_0}{\sqrt{z^2-g^2}}) \right. \nonumber \\
 && \left. \times \sqrt{z^2-g^2}\cos\left
(\frac{zr}{z^2-g^2}\right) \sin \left(\frac{gr}{z^2-g^2}\right) \right. \nonumber \\
 && \left.+  \frac{1}{2}\int_0^g dz 
e^{\frac{ -\left|x_0\right|}{\sqrt{g^2-z^2}}} \sqrt{g^2-z^2}\right. \nonumber \\
 && \left. \times
\cos\left(\frac{zr}{g^2-z^2}\right) \sin \left(\frac{gr}{g^2-z^2}\right) \right].
\end{eqnarray}
As we have mentioned there is a response 
of the fields before the source has acted which reveals violations of causality.
\section{The quantum field theory}
In the next first subsection we compute the propagator 
in the covariant Lorentz gauge in any preferred background. In the second
we compute the commutator function for a purely spacelike background giving
an estimation of the
microcausality violation for spacelike separations near the light cone.
The purely timelike background has been studied in \cite{reyes}. 
\subsection{The propagator in the Lorentz gauge}
Consider the free Lagrangian density (\ref{M-P.LAGRANGIAN}) 
\begin{eqnarray}
\mathcal L&=&-\frac{1}{4}F_{\mu \nu}F^{\mu 
\nu}-   \frac{g}{2} n_{\mu}\epsilon^{\mu\nu \lambda \sigma} A_{\nu}(n 
\cdot \partial)^2   F_{\lambda \sigma}\nonumber \\ &&-\frac{1}{2}
(\partial^{\mu} A_{\mu})^2,
\end{eqnarray}
where we have included a Lorentz gauge fixing term.
We can write
\emph{modulo} total derivatives 
\begin{eqnarray}
\mathcal L=\frac{1}{2} A_{\nu} \left(\eta^{\nu\sigma }
\Box  -2g \epsilon ^{\mu \nu \lambda \sigma} n_{\mu}  (n 
\cdot \partial)^2   \partial_{\lambda}\right)  A_{\sigma},
\end{eqnarray}
where we identify the photon kinetic operator
\begin{eqnarray}
(\Delta^{-1}) ^{\nu \sigma }= \eta^{\nu\sigma }
\Box  -2g \epsilon ^{\mu \nu \lambda \sigma} n_{\mu}  (n 
\cdot \partial)^2   \partial_{\lambda}.  
\end{eqnarray}
We want to find the Feynman propagator by inverting 
the above operator, to this aim
we go to the momentum representation considering 
$A_{\mu}(x)= A_{\mu}(k)e^{-i(k\cdot x)}$ 
to obtain
\begin{eqnarray}
(\Delta^{-1}) ^{\nu \sigma }=-k^2\eta^{\nu \sigma } 
+2ig \epsilon ^{\nu \mu \lambda \sigma} n_{\mu}  (n 
\cdot k)^2 k_{\lambda}.
\end{eqnarray}
The Feynman propagator resulting from the inversion is
\begin{eqnarray}
(\Delta_F(k))_{ \sigma \lambda }&=&\frac{1}{G}
\Big[-k^2\eta_{\sigma \lambda } +2ig (n\cdot k)^2 
\epsilon_{\sigma \alpha \beta \lambda} n^{\alpha} 
k^{\beta}\nonumber \\ &&
- 4g^2 (n\cdot k)^4 \left(n_{\sigma}n_{\lambda}+
 k_{\sigma}k_{\lambda}
\left(\frac{n^ 2}{k^2}\right)\right. \nonumber \\
&&\left. -(n_{\sigma}k_{\lambda}+
n_{\lambda}k_{\sigma})\frac{(n\cdot k)}{k^2} 
\right) \Big],\nonumber \\
\end{eqnarray}
with the pole structure dominated by
\begin{eqnarray}
G=(k^2)^2-4g^2(n \cdot k)^4 ((k\cdot n)^2-k^2n^2).
\end{eqnarray}
  
The case $n=(1, 0,0,0)$ correctly reproduces 
the one calculated in the reference \cite{reyes}
\begin{eqnarray}
(\Delta_F(k)) _{\mu \nu }&=&\frac{1}{((k^{2})^{2} 
-4g^{2}k_{0}^{4}\left| \mathbf{k}%
\right| ^{2})} \Big[ -k^{2}\eta _{\mu \nu } \nonumber \\
&&+2igk_{0}^{2}\epsilon
^{lmr}k_{m}\eta _{l\mu }\eta _{r\nu }  -\frac{4g^{2}k_{0}^{4}}{k^{2}}%
k_{l}k_{r}\delta _{\mu }^{l}\delta _{\nu }^{r} \nonumber \\
&&+
\frac{4g^{2}k_{0}^{4}{\left|
\mathbf{k}\right| }^{2}}{k^{2}}\eta _{0\mu }
\eta _{0\nu }\Big] .
\label{LORGAUGE}
\end{eqnarray}
\subsection{Microcausality}
Consider the commutator of the gauge fields 
\begin{eqnarray}\label{commutator}
[{ A}_{i} (z), {A}_{j} (0)] = i
T_{ij}(-i\partial_0 ,-i\nabla \,)\,D(z) \;,
\end{eqnarray}
where the tensor $T_{ij}(-i\partial_0 ,-i\nabla \,)$ 
is given by the expression (\ref{transv-op}) and
recall from Eq. (\ref{causal-function}) the scalar Green function
\begin{eqnarray}\label{P-J-S}
D(z)=\frac{4\pi}{(2\pi)^4}\oint_{C} d^4k \frac{e^{-ik\cdot 
z}}{(k^2)^2-4g^2(n \cdot k)^4 
 \left((k\cdot n)^2-k^2n^2\right)}.\nonumber \\
\end{eqnarray}
A few observations are in order.
The non locality of the tensor $T_{ij}$ may introduce 
apparent microcausality violations in the commutator (\ref{commutator}) which 
can be bypassed by considering physical fields such as the electric and magnetic fields. 
Therefore, in the following we will consider the 
commutator (\ref{commutator}) involving only physical fields which 
amounts to introduce more derivatives
and to possibly modify the tensor structure of $T_{ij}$. 
We stress that the scalar Green function
the function relevant for the study of causality of the theory remains intact. 

In the following to provide an estimation of microcausality violations
we consider spacelike separations $z^2<0$.
Consider the integral (\ref{P-J-S}) for a purely spacelike four vector $n$
\begin{eqnarray}\label{causal-function2}
D(z)=\frac{4\pi}{(2\pi)^4}\oint_{C} d^4k  \frac{e^{-ik\cdot z}}{
 (k_0^2-\omega_{+}^2)(k_0^2-\omega^2_{-})},
\end{eqnarray}
where the solutions $\omega_{\pm}$ are
given by (\ref{spacelikecase}).
Without loss of generality we will consider $n=(0,0,0,1)$ in which case
the integral (\ref{causal-function2}) takes the form
\begin{eqnarray}\label{space-causal}
 D(z)=\frac{4\pi}{(2\pi)^4}  \oint_{C} d^4k  
 \frac{e^{-i(k_0 z_0- {\bf k}\cdot {\bf z}  )}}{(k^2_0-\omega^2_{+})
 (k_0^2-\omega^2_{-})},
\end{eqnarray}
with  
\begin{eqnarray}
\omega_{\pm}^2=k_{1}^2+k_{2}^2+\widetilde \omega^2_{\pm}(k_{3}) ,
\end{eqnarray}
and the function of $k_3$ given by
\begin{eqnarray}
\widetilde \omega_{\pm}(k_{3})= \sqrt{ k_{3}^2+g^2k_{3}^4 }\pm gk_{3}^2. 
\end{eqnarray}
Unfortunately we cannot simplify by
performing a Lorentz transformation to a frame where 
$z_0=0$, as achieved in the usual case ,
since any finite boost would then generate a zeroth component for $n$ converting
the dispersion relation into a higher order polynomial with imaginary solutions. 

We can still perform a boost in the perpendicular plane  
in order to simplify, namely 
\begin{eqnarray}
k_0 z_0-({\bf k}_{\perp}\cdot {\bf z}_{\perp})  
-k_3 z_3=k'_0  z_0-({\bf k}'_{\perp}\cdot {\bf z}_{\perp})  -k_3 z_3,
\end{eqnarray}
where 
\begin{eqnarray}
k'_0 &=&\gamma(k_0- ({\bf k}_{\perp}\cdot {\bf z}_{\perp}) ),
\\
{\bf k}'_{\perp}&=&{\bf k}_{\perp}+ \Big(\frac{(\gamma-1)}{v^2}
({\bf k}_{\perp}\cdot {\bf z}_{\perp})   -\gamma k_0\Big) {\bf v}.
\end{eqnarray}
With a suitable additional boost for $z$ we can write 
(\ref{space-causal}) as
\begin{eqnarray}\label{integral1}
D(\bar z_0,z_3)=\frac{4\pi}{(2\pi)^4}\oint_{C} d^4k 
 \frac{e^{-i(k_0 \bar z_0-k_3 z_3)}}{(k_0^2-\omega_{+}^2)
 (k_0^2-\omega_{-}^2)},
\end{eqnarray}
where
\begin{eqnarray}
\bar z_0=\sqrt{\frac{z_0^ 2-z^2_{1}-z^2_{2}}{z_0^2}}\,z_0.
\end{eqnarray}
We note that for $ z_0^2 <z^2_{1}+z^2_{2}$ 
we can set $\bar z_0=0$ and therefore the integral (\ref{integral1})
vanishes due to the symmetric contribution of the poles. 

For $z_0^2 \geq z^2_{1}+z^2_{2}$ consider the 
integral in the $k_0$ plane 
\begin{eqnarray}
I(\bar z_0)=\oint_{C} dk_0 
 \frac{ e^{-ik_0 \bar z_0} }{ (k_0^2-\omega_{+}^2)
 (k_0^2-\omega_{-}^2)}.
\end{eqnarray}
Integrating we arrive at
\begin{eqnarray}
&&D(\bar z_0,z_3)=\frac{4\pi}{(2\pi)^3}\int d^3k \;
  e^{ik_3 z_3} \nonumber \\
  && \left(\frac{\sin  \omega_- \bar z_0}{ \omega_- ( 
  \omega_+^2- \omega_-^2)}-\frac{\sin  \omega_+ \bar z_0}
  { {\omega}_+( {\omega}_+^2- {\omega}_-^2)}\right).
\end{eqnarray}

We change to polar coordinates for the 
perpendicular variables $k_1$ and $k_2$ 
such that
\begin{eqnarray}
&& D(\bar z_0,z_3)=\frac{4\pi}{(2\pi)^2}\int_{-\infty}^{\infty}  
dk_3 e^{ik_3 z_3} \int_{0}^{\infty}  \rho \, d\rho \;\nonumber \\
  &&\times
 \left(\frac{\sin  (\omega_- \bar z_0)}{ \omega_- ( 
  \omega_+^2- \omega_-^2)}-\frac{\sin ( \omega_+ \bar z_0)}
  { {\omega}_+( {\omega}_+^2- {\omega}_-^2)}\right), 
\end{eqnarray}
where
\begin{eqnarray}
\omega_{\pm}^2=\rho^2+\widetilde \omega^2_{\pm}(k_{3}) ,
\end{eqnarray}
Given that $ {\omega}_+^2- {\omega}_-^2$ is a function 
of $k_3$ only
the integral in $\rho$ is easily done arriving at
\begin{eqnarray}
&&D(\bar z_0,z_3)=-\frac{4\pi}{(2\pi)^2\bar z_0}\int_{-\infty}^{\infty} dk_3 
 \nonumber \\
 && \times e^{ik_3 z_3}\left(\frac{\cos( \widetilde \omega_+ \bar z_0)
  -\cos (\widetilde \omega_- \bar z_0)}{ \widetilde
  \omega_+^2-\widetilde \omega_-^2}\right) ,
\end{eqnarray}
where we are neglected a fast oscillating part.

Let us rewrite 
\begin{eqnarray}
  &&D(\bar z_0,z_3)=-\frac{1}{8\pi\bar z_0}
  \int_{-\infty}^{\infty} dk_3 e^{ik_3 z_3}\nonumber \\
 && \times
  \left(  \frac{e^{i\widetilde \omega_+ 
  \bar z_0}+e^{-i\widetilde \omega_+
   \bar z_0}-e^{i\widetilde \omega_{-} 
   \bar z_0}-e^{-i\widetilde 
   \omega_{-} \bar z_0}   }{gk_3^2 
   \sqrt{ k_3^2+g^2k_3^4 }   }\right),
\end{eqnarray}
and let us define
\begin{eqnarray}
D(\bar z_0,z_3)&=&-\frac{1}{8\pi\bar z_0}(I_{1+}(\bar z_0)
\nonumber \\
&&+I_{1-}(\bar z_0)-I_{2+}(\bar z_0)-I_{2-}(\bar z_0)).  
\end{eqnarray}
The integrals that shall be computed have the form
\begin{eqnarray}
I_{a\lambda}(\bar z_0)=\int_{-\infty}^{\infty} dk_3 f(k_3)
   e^{i \Phi_{a\lambda}(k_3)},
\end{eqnarray}
with the phases being
\begin{eqnarray}
 \Phi_{a\lambda}(k_3)=k_3 z_3+ \lambda \bar z_0 
 (\sqrt{ k_{3}^2+g^2k_{3}^4 }-(-1)^a gk_{3}^2),
\end{eqnarray}
for $a=1,2$ and the function
\begin{eqnarray}
f(k_3)=   \frac{1}{gk_3^2 \sqrt{ k_3^2+g^2k_3^4}}.
\end{eqnarray}
We use the stationary method 
to approximate the above integrals as 
\begin{eqnarray}
I_{a\lambda}(\bar z_0)=f(\bar k_3) e^{i \Phi_{1,2\lambda}
(\bar k_3)} \int_{-\infty}^{\infty} dk_3 
   e^{\frac{i}{2} \Phi''(\bar k_3)( k_3-\bar k_3)^2},  
\end{eqnarray}
for a stationary point $\bar k_3$.

Notice that for $\bar z_0<0$ there are no stationary points and for
 $\bar z_0>0$ 
we have stationary points only 
for $I_{1+}(\bar z_0)$ and $I_{1-}(\bar z_0)$ which are respectively 
$\bar k_3=\frac{-(z_3-\bar z_0)}{2 g \bar z_0}$ and
$\bar k_3=\frac{(z_3-\bar z_0)}{2 g \bar z_0}$ near the light cone. 
After a straightforward calculation we arrive at
\begin{eqnarray}
D(\bar z_0,z_3)&=&  \sqrt{\frac{2}{\pi}} \frac{(g\bar z_0)^{3/2}}{(  z_3-\bar z_0)^3} \nonumber \\
&& \times
\left[\cos \left( \frac{(  z_3-\bar z_0)^2}{4g\bar z_0} \right)
+ \sin \left( \frac{(  z_3-\bar z_0)^2}{4g\bar z_0} \right) \right],\nonumber \\
\end{eqnarray}
where we have used the integrals 
\begin{eqnarray}
\int_{-\infty}^{\infty} dx e^{\pm ig\bar z_0 x^2} =
\sqrt{\frac{\pi}{2g \bar z_0}}(1\pm i).
\end{eqnarray}
We observe that the microcausality violation is suppressed by a power $g^{3/2}$.
\section{Conclusions}
Motivated by astrophysical observational tests 
and their strong limits on the Myers and Pospelov
parameter we have considered the 
possibility to extend the existing treatment for  
spacelike and lightlike privileged backgrounds.
Phenomenological studies based on anisotropies 
introduced by the spacelike theory
are out of the aim of the present work but seems 
a natural transition. Some works 
have already been started \cite{spacelike}.
The major interest in this work focuses on the consistency
of the effective theories that introduce higher dimensional operators.
Therefore it shall be considered preliminary work 
before any phenomenological computation. 

It is well known that higher derivative theories 
can lead to instabilities, negative norm states,
and nonunitarity problems to mention some.
We have found that the purely spacelike theory is 
free of the previous issues which strengthen
the idea to extend the existing searches for 
Planck scale phenomena maintaining dimension-5 operators.
It also opens the concrete possibility to quantize the theory 
without the introduction of any cut-off function
as required in the timelike case.

To summarize, we have obtained the covariant dispersion relation 
for the propagation of photons in all privileged frames.
The timelike, spacelike and lightlike 
preferred backgrounds have been treated separately for the 
study of photon kinematic properties. In addition,
asymptotes properties of the group velocity, had given us a
well idea about causality and stability within the classical theory.
For the timelike case we 
have found runaway solutions above certain values of 
momenta which suggest the introduction of 
a cut-off function. 
The quantization of a model and the implementation of such cut-off 
function has been carried out in \cite{reyes}.
For the lightlike case we have found additional degrees 
of freedom with some of them being non analytical.
Using the criteria of analyticity we have regarded the 
lightlike theory as a genuine higher-derivative theory.
As further work it would be interesting to identify 
possible negative energies 
and to analyze in detail potential negative norm states 
arising in the theory. 
Given the similarity with the Chern Simons modification 
it would be interesting also 
to study possible extended symmetries and on shell field 
redefinitions \cite{Ralf2}.
We have found that the purely spacelike effective field theory
is stable and
causal while microcausality is highly suppressed. It is therefore appropriate to be quantized.
This gives the spacelike Myers and Pospelov theory the possibility
to play a role in future searches for Lorentz violation.
\section*{Acknowledgments}
I am indebted to H. A. Morales Tecotl, R. Lehnert, 
M. Cambiaso, J. Gamboa and F. Mendez for valuable 
comments on this work and encouragement. This work was 
partially supported by Mexico's National
Council of Science and Technology CONACyT-SEP under
Grant No. 51132F
and by Postdoctoral Project DICyT of Usach.

\end{document}